# Magnetocaloric properties of an inhomogeneous magnetic thin film of 7.6 nm La$_{0.7}$Sr$_{0.3}$MnO$_3$ grown on SrTiO$_3$.

Navid Mottaghi[1], Robbyn B. Trappen[1], Saeed Yousefi[1], Mohindar S. Seehra[1], Mikel B. Holcomb[1]

[1]Department of Physics & Astronomy, West Virginia University, Morgantown, WV 26506, USA.

## Abstract

Magnetocaloric properties of an inhomogeneous magnetic system of a 7.6 nm La$_{0.7}$Sr$_{0.3}$MnO$_3$ consisting of superparamagnetic (SPM) with blocking temperature ($T_B$ =240 K) and ferromagnetic (FM) phases ($T_C$ = 290 K) is studied by dc magnetization measurements. Isothermal magnetization versus applied magnetic field is carried out from 100 K to 320 K in magnetic fields up to 4 kOe to determine changes in the magnetic entropy ($-\Delta S_M$) and the relative cooling power (RCP). Due to the co-existence of SPM and FM phases, there are two peaks in the temperature dependence of $-\Delta S_M$ in different applied magnetic fields from 1.3 kOe to 4 kOe. The peaks are at 220 K and 270 K which are close to $T_B$ and $T_C$ of the film. The highest RCP occurs at 270 K (which is in $T_B < T < T_C$) in $H$ = 4 kOe with the value of 0.19 (J/kg K). The $-\Delta S_M$ vs $T$ data are fit to the exponent power law, $-\Delta S_M = aH^n$ where it shows good fits for the whole measured temperature range. This analysis reveals a deviation of $n$ from $n$ = 2/3 which is likely due to the presence of SPM spin clusters in the dead layer for $T < T_C$. Results show that the thin film of La$_{0.7}$Sr$_{0.3}$MnO$_3$ can be a good candidate for magnetic refrigeration devices with multiple RCP peaks in low and high temperatures.

I. Introduction

Magnetic refrigerators (MRs) work upon the magnetocaloric effect (MCE). MCE occurs when the temperature of a magnetic material changes as it is exposed to an applied magnetic field under the adiabatic condition. MR can be a good substitution for conventional vapor compression refrigerators as they have many advantages over conventional gas compression (CGC) technology. The energy consumption of MRs is 20-30% less than CGS refrigerators and the efficiency can be up to 30-60% of a Carnot cycle [1, 2]. The heart of MR is the working magnetic material which plays important roles in MR technology. Choosing a magnetic material with good MCE effects and a room temperature magnetic transition paves the way to make room temperature MR devices [8]. them depends on many factors, e.g., the magnetic material particle size [3], magnetic phase transition temperature [4], the strength of external applied magnetic field [5], presence of different magnetic phases [6], structural phase transitions [7], etc.

Among MCE materials, manganites have been considered as good candidates due to their good MCE efficiency and a tunable MCE peak in different temperature regions up to room temperature [3, 5, 8]. For example, the La$_{0.7}$Sr$_{0.3}$MnO$_3$ (LSMO) structure, has the highest Curie temperature ($T_C$) around 370 K in thin films [9] and nanoparticles [5] which makes this material a suitable candidate for room temperature MRs. Optimizing an MR material with the properties discussed above is one of the optimum goals of MCE studies. Also desired is the single and sharp magnetic entropy change at a transition temperature. However, the existence of multiple magnetic phases in a magnetic material creates different magnetic entropy changes in the vicinity of phase transition of each phase or a broad distribution of change in entropy over a wide temperature range[10, 11].

The magnetic inhomogeneity due to the presence of multiple magnetic phases in a magnetic material has been observed in an LSMO thin film [12] and many other magnetic systems [13]. Recently we studied bulk magnetometry of a 7.6 nm $La_{0.7}Sr_{0.3}MnO_3$ (LSMO) grown on $SrTiO_3$ (STO) substrate with the transition temperature of ($T_C$) around 290K to understand the so-called dead layers/regions. We have reported that these layers/regions are formed by randomly distributed spin clusters[14]. These clusters cause the bifurcation between in zero-field-cooled (ZFC) and field-cooled (FC) cycles in measurements of the temperature variation of magnetization (M-T). This bifurcation temperature was interpreted as a blocking temperature ($T_B$). The $T_B$ was measured from M-T data at 50 Oe applied field and ZFC hysteresis loops. These measurements supported a $T_B$ around 240 K [12].

Also, it was shown these clusters were responsible for the reduction of saturation magnetization ($M_S$) with the domain diameter and the thickness of 90 nm and 1.4 nm, respectively[14]. Moreover, we could show these clusters are acting like superparamagnetic domains embedded in the ferromagnetic medium of the sample, making it an inhomogeneous magnetic system. The co-existence of magnetic phases has been proven by the presence of inverted hysteresis loops (IHLs) [12].

Here the magnetocaloric properties of a 7.6 nm inhomogeneous magnetic LSMO thin film on an STO substrate are studied by using dc magnetization measurements. The change of magnetic entropy is observed around $T_C$ and $T_B$. The former is due to FM-PM magnetic phase transition and the latter is originated from the presence of spin clusters. The change of induced specific heat capacity is also calculated where significant effects of SPM clusters has been observed around $T_B$. Moreover, relative cooling power (RCP) is calculated around $T_B$ and $T_C$ to understand the role of an inhomogeneous magnetic system in the change of magnetic entropy.

## II. Experiments

The sample is an LSMO thin film with a thickness of 7.6 nm grown on a SrTiO3 (001) substrate. It was grown by pulsed laser deposition (PLD) using a commercial Neocera PLD system with 248 nm KrF excimer laser [12]. The detailed growth conditions have been discussed in our previous publications [12, 14]. Magnetization (M) measurements were performed as a function of the applied magnetic field (H) and temperature (T) using a Quantum Design physical property measurement system (PPMS-9T). M vs. T measurements were recorded in zero-fled-cooled (ZFC) and field-cooled (FC) cycles from 5K to 320K [12]. The Demagnetization process was performed by heating up the sample above the magnetic ordering temperature. For each measurement, the magnet coil was first demagnetized in the oscillating mode so that the residual H is reduced to

<2 Oe. In the ZFC cycle, after demagnetizing the sample, it was cooled down to 5 K in $H$ = 0 Oe. ZFC $M$ vs. $H$ measurements were obtained in the range of $0 \leq H \leq 4$kOe.

III. Magnetocaloric properties

The isothermal $M$ versus $H$ data presented in Fig. 1 is used to determine the magnetocaloric properties of the sample. Based on Maxwell relations the isothermal magnetic entropy change, $\Delta S_M$ ($T$, $H$), as a function of $H$ is given by

$$\Delta S_M(T,H) = S_M(T,H) - S_M(T,0) = \int_0^H \left(\frac{\partial M(T,H')}{\partial T}\right)_{H'} dH', \tag{1}$$

$M$ is measured at discrete $H$ and temperature intervals, and $\Delta S_M(T,H)$ can be approximately calculated by the following equation[15]:

$$\Delta S_M(T,H) = \sum_i \frac{M_{i+1}(T_{i+1},H) - M_i(T_i,H)}{T_{i+1} - T_i} \Delta H. \tag{2}$$

Therefore, the magnetic entropy change at temperature $T$ is the summation of two isothermal magnetization curves at $T$ and $T + \Delta T$ divided by the $\Delta T$ (the temperature difference between two isothermal magnetizations) in Fig. 1. For example: as is typical [5] the value of $\Delta S_M$ is negative in the whole temperature range. There are two peaks at 220K and 270K which are close to $T_B$ and $T_C$ of the sample, respectively. As we discussed in the introduction section, this sample is magnetically inhomogeneous with two magnetic phases. The spin clusters act like SPM phases having a blocking temperature of $T_B$ = 240K and the FM phase has the transition temperature of $T_C$ = 290K. This magnetic inhomogeneity has changed the -$\Delta S_M$ peak spectrum in the whole temperature range. There is not a single sharp peak at the FM-PM transition temperature. From Fig.2 one can see that there are two broad peaks at 220 K and 270 K where -$\Delta S_M$ has the highest value at the latter temperature. The value of -$\Delta S_M$ at peaks increases as the applied magnetic field increases. For example, at 270K, the value of -$\Delta S_M$ is 0.061 (J/kg K) at 1.3 kOe and it increases to 0.19 (J/kg K) at 4 kOe.

Normally in a homogeneous magnetic system, there is only one peak in $\Delta S_M$ which is around the transition temperature of the magnetic phase. For example, in magnetically homogeneous LSMO nanoparticles the peak occurs at around $T_C$ [16]. On the other hand, it might not be a single and sharp peak at the temperature variation of $\Delta S_M$. This is true when a magnetic material is magnetically inhomogeneous[17] or the surface to volume ratio of a sample is significantly large[18]. Then the working temperature ($\Delta T_{\text{FWHM}}$) defined as the full width at half maximum of magnetic entropy change peak is broadened or there can be maximums in the $\Delta S_M(T,H)$ graph.

Field variation of $-\Delta S_M$ is studied to quantify the magnetic state of the sample. To understand this behavior, the data in Fig. 3-a is fit to an exponent power law: $-\Delta S_M = aH^n$ where 'a' is a constant and the exponent 'n' depends on the magnetic state of the sample[4, 19]. Determination of the 'n' exponent allows us to find suitable theoretical models to explain MCE in the system. It is predicted that the value of 'n' is quadratic in the Curie law above $T_C$ and comparing with molecular field theory this value is predicted to be 2/3 [4]. It was reported that the value of 'n' in polycrystalline and nanocrystalline manganites at the Curie temperature is below 1 and in the ranges of $T<T_C$ ($T>T_C$) 'n' reaches to 1 (=2), respectively [19]. Our fit of the data to this Eq. for several $T \leq T_C$ in Fig. 3-b shows $n \sim 1$ for $T<T_C$ with the magnitude of $n$ increasing for $T>T_C$. This deviation of n from $n = 2/3$ is likely due to the presence of SPM spin clusters in the dead layer for $T<T_C$. At $T_B$, the value of $n$ is around 0.83 which is the signature of SPM phase in magnetocaloric studies [19]. The larger magnitudes of n for $T>T_C$ is due to the Curie-Weiss variation of the magnetization in this regime.

$\Delta S_M(T, H)$ can be obtained from the field dependence of the specific heat through the subsequent integration.

$$\Delta S_M(T, H) = \int_0^T \frac{C_P(T, H) - C_P(T, 0)}{T} dT, \qquad (3)$$

where $C_P(T, H)$ and $C_P(T, 0)$ are the measured heat capacity values in an applied field $H$ and zero applied field, respectively. From Eq. 3, one can calculate the change of induced specific heat by an applied $H$ as

$$\Delta C_P(T, H) = C_P(T, H) - C_P(T, 0) = T \frac{\partial(\Delta S_M(T, H))}{\partial T}. \qquad (4)$$

Using Eq. (4) temperature variation of $\Delta C_P$ of the sample is calculated at different applied fields and it is shown in Fig. 4. There are anomalies in all curves around $T_B$ of the sample. The value of $\Delta C_P$ changes quickly from positive to negative around the blocking temperature and it decreases with increasing the temperature. Around $T_C$, $\Delta C_P$ has a value of 1.19 (J/kg K) measured at 4 kOe applied field. Usually, in a homogeneous FM magnetic material, it is expected to see a cusp at $T_C$ where the magnetic material changes its magnetic state from FM to PM [16]. However, this is not the case here since this sample is magnetically inhomogeneous and it has a short range FM phase above $T_C$ [12] and the cusp doesn't appear in $\Delta C_P$ data. The presence of SPM phase in this sample changes the trend of heat capacity in the temperature range $T<T_C$.

To quantify the efficiency for refrigeration the relative cooling power (RCP) is calculated. RCP is the amount of heat transfer between the cold and the hot reservoir in a complete refrigerator cycle [20]. This parameter is based on the change of the magnetic entropy as

$$RCP(S) = -\Delta S_M(max) \times \Delta T_{FWHM}, \qquad (5)$$

where $\Delta S_M(max)$ is the maximum of the entropy change and $\Delta T_{FWHM}$ is the working temperature as defined before. In order to calculate $\Delta T_{FWHM}$ it was required to fit four Gaussian functions around $T_B$ and $T_C$ to $\Delta S_M$ graphs in Fig. 2. The center of the Gaussian peaks are at 220, 250, 270 and 290 K, which are located at $T_B<T<T_C$ temperature region. The fit results are presented in the supplementary material in Fig. S1. Figure 5 shows the RCP value of the 7.6 nm LSMO/STO sample. Due to magnetic inhomogeneity, the maximum RCP value occurs at $T_B$< 270 K <$T_C$ which increases from 0.75 (J Kg$^{-1}$) to 3.75 (J kg$^{-1}$) at 4 kOe applied field which is comparable with low field measurements [5].

## IV. Conclusions

In summary, the magnetocaloric properties of an inhomogeneous magnetic system of a 7.6 nm LSMOfilm on STO with $T_B$ =240 K and $T_C$ = 290 K are studied. The co-existence of SPM and FM phases in this sample creates a considerable MCE around its $T_B$ and $T_C$ temperatures. For the small field of 4 kOe, the peak values of $-\Delta S_M$ is 0.13 (J/kg K) and 0.19 (J/kg K) at 220 K and 270 K, respectively. More detailed inspection of the magnetic field dependence of $-\Delta S_M$ at different temperatures was explored by fitting to an exponent power law: $\Delta S_M = aH^n$ where $n$ was found to be 0.83 at $T_B$ which is the signature of an SPM phase in magnetocaloric studies. Field dependence of specific heat capacity shows a clear cusp at $T_B$. However, there no cusp is observed at around $T_C$ due to the presence of the short-range FM phase at the temperature range of $T>T_C$ where $\Delta S_M \neq 0$ and the system goes to another ordered magnetic phase. The so-called relative cooling power is calculated around $T_B$ and $T_C$ and it has the highest amount at 270 K. Finally, the role of magnetic inhomogeneity in the thermodynamic properties of the LSMO/STO thin film is discussed here. On the positive side, one can take advantage of having different RCPs in a wide temperature range in a single magnetic material. But on the other side, this can reduce the RCP of magnetic thin films which can be problematic to industrial applications.

**Acknowledgments:** We thank the assistance of Dr. Qiang Wang who is helping us to maintain the physical property measurement system (PPMS) machine. We acknowledge funding support from NSF (DMR-1608656) for growth and optimization and DOE (DE-SC0016176) for the magnetic characterization of our films. This work has been included in our sample database made possible by NASA WV EPSCoR Grant # NNX15AK74A.

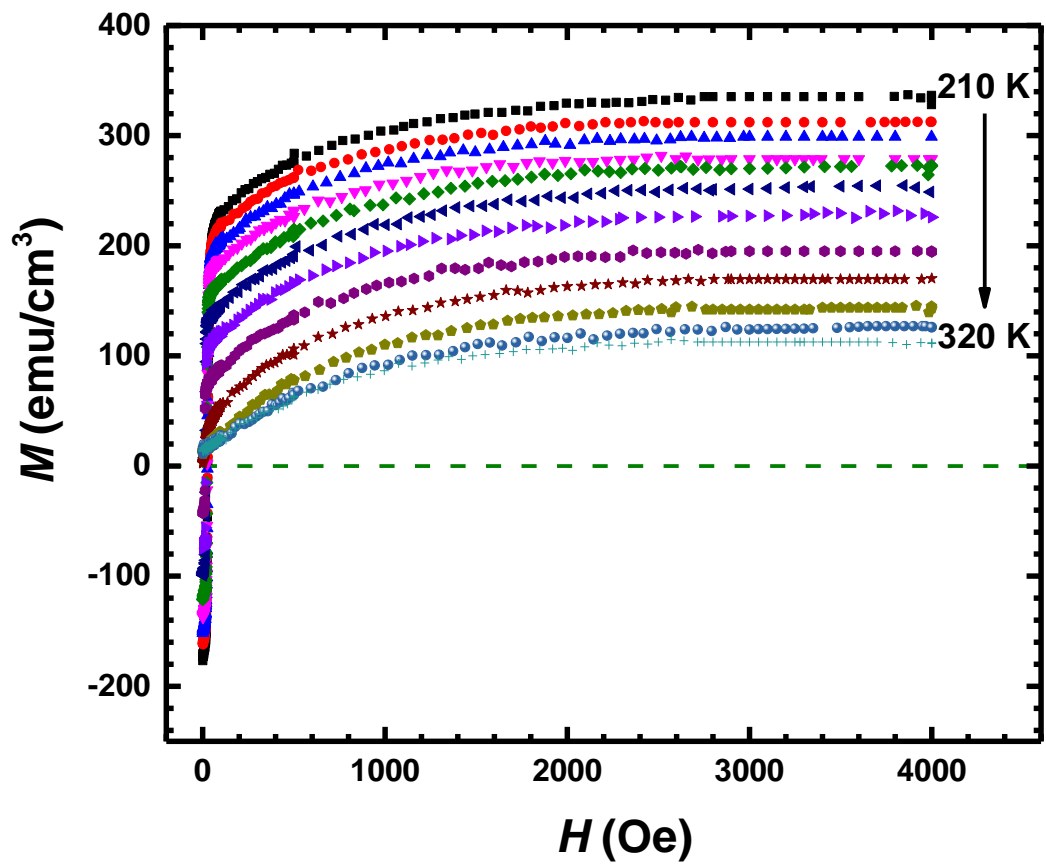

FIG. 1. Selected isothermal *M* versus *H* curves at selected temperatures from 210 K to 320 K.

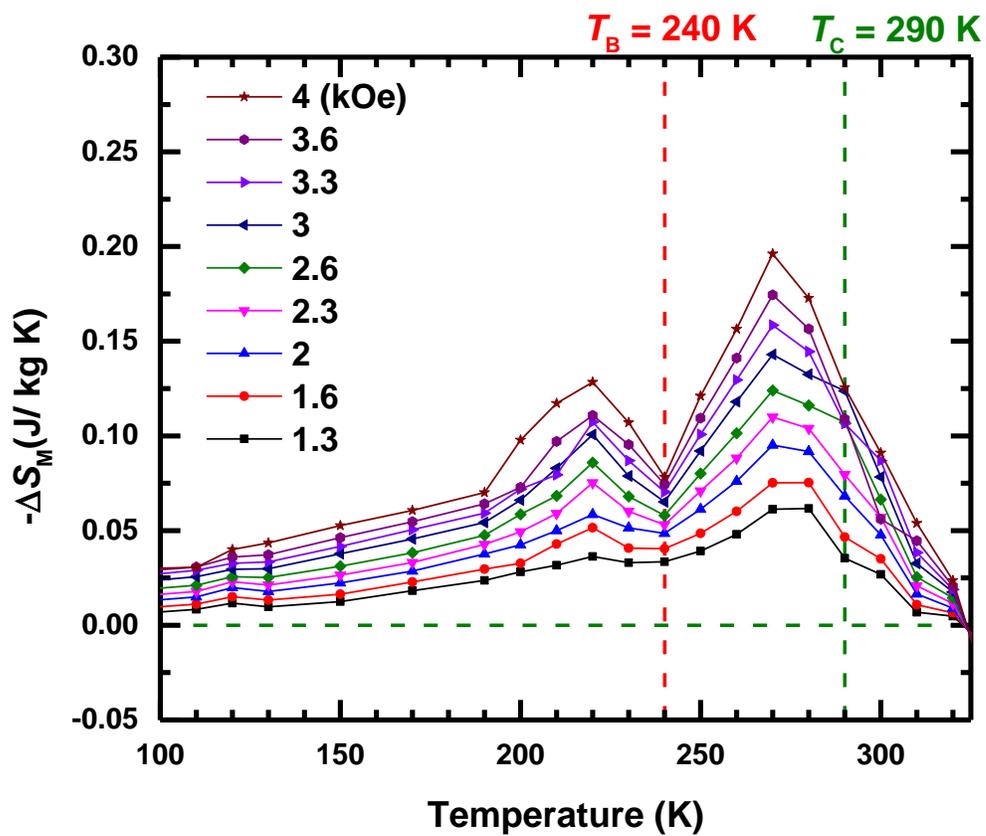

FIG. 2. Magnetic entropy of the sample versus temperature at different applied magnetic fields.

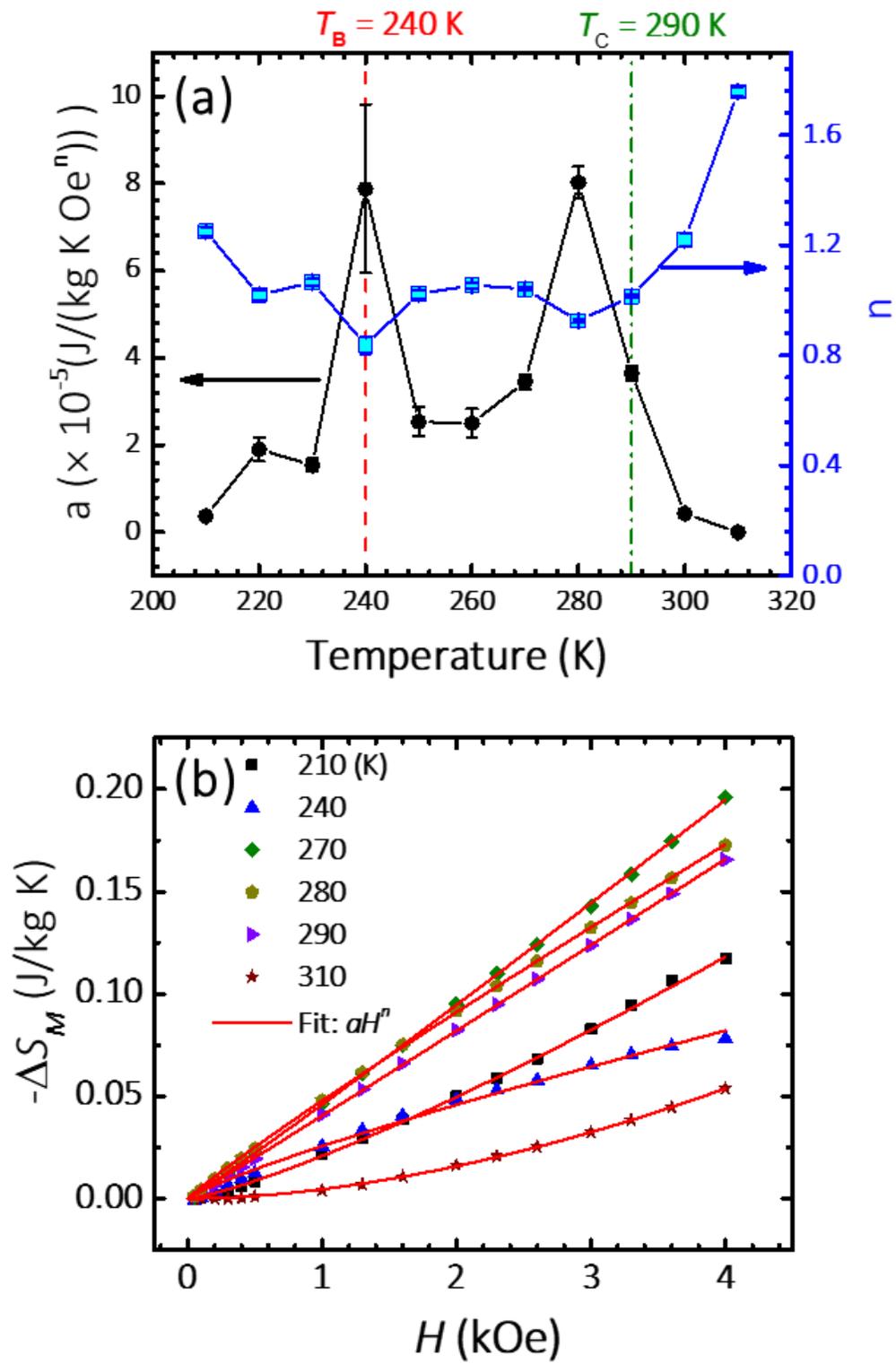

Fig. 3. (a) Magnetic field dependence of the change in magnetic entropy at different temperatures. (b) Temperature dependence of 'a' and the exponent 'n' versus temperature.

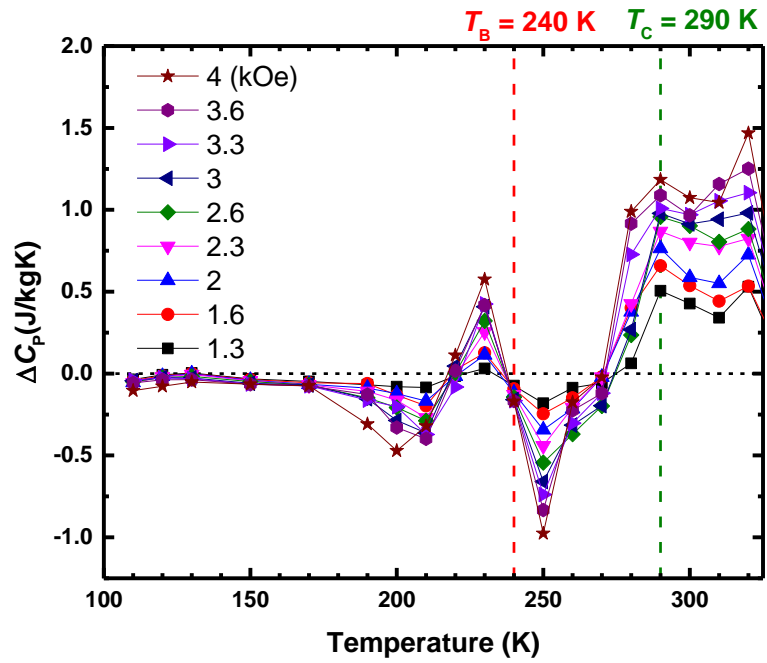

FIG. 4. Field dependence of specific heat of the sample as a function of temperature.

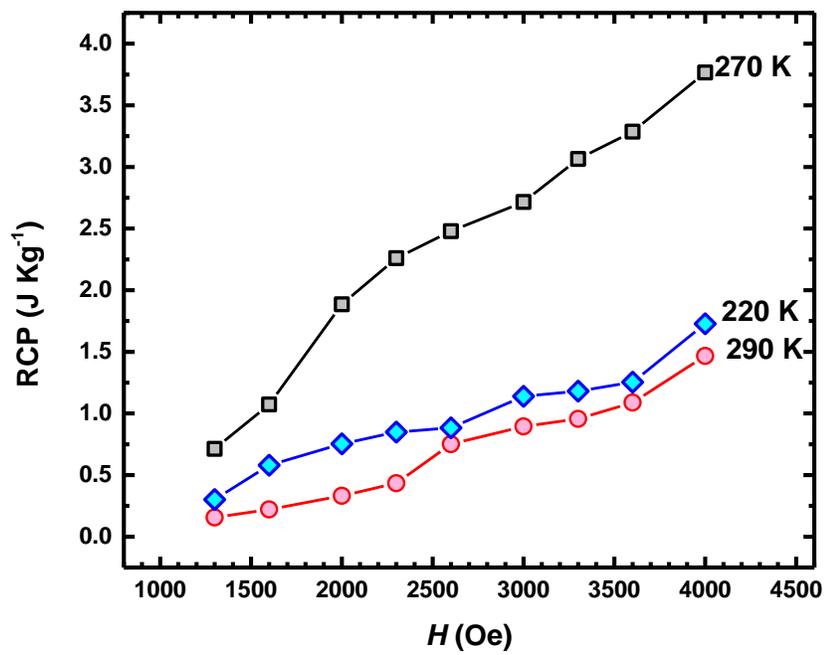

FIG. 5. Relative cooling power (RCP) of 7.6 nm LSMO/STO the sample measured at 220, 270 and 290 K.